\documentclass[english,aps,prd,twocolumn,showkeys]{revtex4-1}
\usepackage[T1]{fontenc}
\usepackage[latin9]{inputenc}
\usepackage{color}
\usepackage{amsmath}
\usepackage{graphicx}

\makeatletter

\@ifundefined{textcolor}{}
{%
 \definecolor{BLACK}{gray}{0}
 \definecolor{WHITE}{gray}{1}
 \definecolor{RED}{rgb}{1,0,0}
 \definecolor{GREEN}{rgb}{0,1,0}
 \definecolor{BLUE}{rgb}{0,0,1}
 \definecolor{CYAN}{cmyk}{1,0,0,0}
 \definecolor{MAGENTA}{cmyk}{0,1,0,0}
 \definecolor{YELLOW}{cmyk}{0,0,1,0}
}

\def\gs{\mathrel{
   \rlap{\raise 0.511ex \hbox{$>$}}{\lower 0.511ex \hbox{$\sim$}}}}
\def\ls{\mathrel{
   \rlap{\raise 0.511ex \hbox{$<$}}{\lower 0.511ex \hbox{$\sim$}}}}
\makeatother

\usepackage{babel}
\begin{document}

\title{Robustness of Neutrino Mass Matrix Predictions}

\author{Werner Rodejohann$^a$ and Xun-Jie Xu$^{a,b}$ }

\affiliation{$^a$Max-Planck-Institut f\"{u}r Kernphysik, Postfach 103980, D-69029
Heidelberg, Germany\\
$^b$Institute of Modern Physics and Center for High Energy Physics,
Tsinghua University, Beijing 100084, China}
\begin{abstract}
We investigate the stability of neutrino mass matrix predictions on
important and currently unknown observables. Those are the octant 
of $\theta_{23}$, the sign of $\sin\delta$ and the neutrino mass
ordering. Determining those unknowns is expected to be useful in order to distinguish neutrino mass models. Therefore it may be interesting to know how robust the predictions of a mass matrix for the octant of $\theta_{23}$ or the neutrino mass
ordering are. 
By applying general multiplicative perturbations we explicitly quantify how probable it is 
that a perturbed mass matrix predicts an octant of $\theta_{23}$ 
different from the original mass matrix, or even a neutrino mass ordering different from 
the original one. 
Both the general case and an explicit flavor symmetry model are studied. 
We give the probabilities as a function of the smallest neutrino mass, showing that for values 
exceeding 0.1 eV the chance to switch the prediction quickly approaches 50\,\%. 

\end{abstract}

\maketitle
\def\headif{\iftrue}

\headif

\section{Introduction}
\noindent
In recent years a consistent picture of lepton mixing has emerged \cite{PDG}, with several  parameters being determined with increasing precision (for a recent global fit  of all existing data, see Ref.\ \cite{fit}). A remarkable pattern has emerged, with 
one close-to-maximal mixing angle, one large and one small mixing angle, the latter being of the order of the largest quark mixing angle.  While the overall picture of the leptonic mixing matrix is clear, comparable precision with respect to the quark sector is still lacking, but future experiments and facilities exist that will improve the errors on the parameters by  remarkable amounts, see e.g.\ \cite{Kettell:2013eos}. Of particular interest in neutrino physics are the octant of the atmospheric neutrino mixing angle $\theta_{23}$ and of course $\sin \delta$, the parameter governing leptonic CP violation. The mass ordering and the value of the smallest neutrino mass are also unknown 
(while not yet determined, we will assume here that neutrinos are Majorana particles). 

The astonishing disparity between lepton and quark mixing has lead to huge efforts in flavor symmetry model building \cite{Altarelli:2010gt,Ishimori:2010au,Feruglio:2015jfa}. Many neutrino mixing schemes have been proposed (see e.g.\ \cite{Albright:2010ap}), and many 
models exist that can generate these schemes. The question is now of course to distinguish the various models or scenarios and identify the correct one. One could expect that the determination of the unknown neutrino parameters, in particular the sign of $\sin \delta$, the octant of $\theta_{23}$ or the mass ordering will be crucial. In this paper we analyze how robust these parameters are with respect to perturbations of the mass matrix. 
Perturbations to a mass matrix are expected to be present because of various reasons, e.g.\ renormalization effects including thresholds, misalignment of the vacuum expectation values of the flavons which are crucial in flavor symmetry models, non-canonical kinetic terms, higher-dimensional operators, etc. By quantifying how probable it is that a perturbed mass matrix changes its predictions for a currently unknown neutrino parameter, one can estimate how robust the predictions are. 
Analyzing this issue is the purpose of the present paper. 
As the probability to change the predictions depends strongly on the neutrino mass scale, this is especially crucial for sizable neutrino masses, with the extreme case being quasi-degenerate neutrino masses.

Our procedure is as follows: we start with a large set of mass matrices that are allowed 
according to current global fits, but have a certain property that is of interest to us, 
say, $\theta_{23}<\pi/4$. Then we multiplicatively 
perturb the mass matrices in a general way, and check 
how many percent of the resulting mass matrices
 change the property of interest, i.e.\ predict $\theta_{23}>\pi/4$ after 
perturbation. This percentage is a function of the smallest neutrino mass. 
We demonstrate that, as one may expect, the percentage increases strongly 
with the smallest neutrino mass and is in general larger for the 
inverted ordering than for the normal one. The solar neutrino mixing angle is 
subject to the largest instability among the mixing angles, the CP phase $\delta$ as well. 
The sign of $\sin \delta$ is more likely to change than the octant of $\theta_{23}$. 
For values of the 
smallest neutrino mass around 0.1 eV and larger, even the mass ordering can 
change from normal to inverted. In general, for neutrino masses larger than 0.1 eV 
the probability to change a prediction quickly approaches 50\,\%. 
While intuitively many findings are expected, there has never been a quantitative study addressing these issues. 
We also analyze an explicit model based on $A_4$, in which correlations 
among the mass matrix parameters are present. We find qualitatively similar results. 
This demonstrates the challenge to distinguish neutrino mass models or scenarios 
unless corrections are taken into account.

The paper is build up as follows: in Section \ref{sec:gen} we present the procedure and then  discuss the perturbation of a general mass matrix. Section 
\ref{sec:a4}  deals with a specific $A_4$ model, before we conclude in Section \ref{sec:concl}.

\section{\label{sec:gen}Perturbation of a general mass matrix}

\subsection{Method}

\noindent
Let us start with a zeroth-order neutrino mass matrix $M_{0}$, constructed by
\[
M_{0}=U_{0}\,\textrm{diag}(m_{1},m_{2},m_{3})\,U_{0}^{T}\, ,
\]
where $U_{0}$, parametrized as usual \cite{PDG}, includes the Majorana phases and $(m_{1},m_{2},m_{3})$
can be determined by the mass-squared differences (following the definition of 
Ref.\ \cite{fit}) $\delta m^{2}=m_{2}^{2}-m_{1}^{2}$, 
$\Delta m^{2}=m_{3}^{2}-(m_{1}^{2}+m_{2}^{2})/2$.  
The lightest neutrino mass is $m_1$ for the normal mass ordering, 
$m_3$ for the inverted one.

We consider now a general multiplicative 
perturbation to the individual mass matrix entries: 
\begin{equation}
(M_{0})_{\alpha\beta}\rightarrow M_{\alpha\beta}=(M_{0})_{\alpha\beta}\, (1+\epsilon_{\alpha\beta})\,,\label{eq:0302}
\end{equation}
where $\epsilon_{\alpha\beta} = \epsilon_{\beta\alpha} $ are six small complex numbers. 
Note that multiplicative perturbations are conservative, one could also add 
terms $\epsilon_{\alpha \beta} \, M_{0}^{\rm max}$ to each entry, i.e.\ corrections proportional 
to the largest entry in the mass matrix. Such additive corrections are expected 
to give qualitatively similar perturbations as the ones we will derive here. However, 
they will be at least as sizable as the multiplicative ones under study, as they 
influence small entries of the mass matrix more significantly (note that with 
multiplicative corrections texture zeros and the associated correlations they 
introduce are not significantly changed). In addition, 
often and extensively studied corrections from the charged lepton sector 
could be included as well. We have nothing new to add to this aspect, and in addition 
those correction are model-dependent, and furthermore independent on neutrino 
mass and ordering. We stick in the present paper to the conservative case of  
multiplicative corrections to mass matrices and the analysis thereof.

One has several possibilities to choose the initial parameters \footnote{Let us note 
in this respect that corrections are not necessarily bad, as a given model could 
have a prediction incompatible with data, and corrections lead to agreement with data.}. We decided to 
choose in $M_0$ the mixing angles and mass-squared differences 
($\theta_{13}^0$, $\theta_{12}^0$, $\theta_{23}^0$ and $(\delta m^{2})^0$, 
$(\Delta m^{2})^0$) randomly within their current 3$\sigma$ confidence intervals, 
while for both Dirac and Majorana CP phases, we randomly generate 
them in $[0,2\pi]$. 
We will however be interested in a certain property, say $\theta_{23}^0<\pi/4$. 
Therefore, this condition is imposed on $M_0$. 
After $M_{0}$ is constructed, we randomly generate the six $\epsilon_{\alpha \beta}$
with $\sum |\epsilon|<0.2$ and $\sum|\epsilon|^{2}>0.01^{2}$. 
We require that $M$ after perturbation (having mixing angles $\theta_{13}$, 
$\theta_{12}$, $\theta_{23}$ and mass-squared differences $\delta m^{2}$, $\Delta m^{2}$) 
is still compatible with 
current data within $3\sigma$. We are interested in the percentage of successful neutrino mass matrices $M$ that are within $3\sigma$, but have went from $\theta_{23}^0 <\pi/4$ to $\theta_{23}>\pi/4$. Put another way, we obtain the probability for the perturbed mass matrix 
to change the characteristic prediction we are interested in. The same procedure is 
performed for the sign of $\sin \delta$ and for the mass ordering. 
We are interested in how the results depend on the smallest neutrino mass. 
To make robust statements, we want 10000 successful mass matrices for each value of 
the smallest mass. Hence, the numerical analysis is quite CPU-intensive in particular 
for neutrino masses near and above 0.1 eV. 

A comment on the upper and lower limits on the $\epsilon$ is in order: 
a compromise between a ``reasonable'' percentage of successful mass matrices after 
perturbation on the one hand, and guaranteed perturbations to the mixing parameters 
on the other hand,  needs to be found. A lower limit on the perturbations is needed because if we 
have no lower limit the vast majority of successful mass matrices is essentially 
identical to the original ones. Larger upper limits on the $\epsilon_{\alpha\beta}$ than the ones 
we use increase the CPU-time for sizable neutrino mass significantly. The limits on the 
$\epsilon_{\alpha\beta} $ might be interpreted similarly to the model analysis in 
Section \ref{sec:a4}, namely as VEV misalignment in a flavor symmetry 
model of order a few percent. With typically 3 to 5 VEVs 
playing a role, see Eqs.\ (\ref{eq:0322-03},\,\ref{eq:0322-3}), 
the upper limit on the sum of $\sum |\epsilon|<0.2$ could be understood. 
We would like to avoid too much cancellations in the $\epsilon$, 
hence a lower limit should be present. 
Threshold effects with RG corrections might be another interpretation of the $\epsilon$. 
We prefer however to stay here as model-independent as possible. In any interpretation of the 
$\epsilon$, a given model might induce a correlation between them. This is realized in 
the model that is studied in Sec.\ \ref{sec:a4}.

Anyway, we have checked that for small neutrino masses, where the analysis takes still reasonable 
CPU-time, the results do hardly depend on the precise values of the upper and lower limits 
of the $\epsilon$, up to longer CPU-time for larger upper limits. 
With this check we gained confidence in our choice of limits.

A few words on generating the events: the most straightforward method to realize it 
is simply "generate-and-reject", which means to generate enough events without the 
constraint and then reject those which violate our constraints. 
This is of low efficiency especially for large values of the smallest neutrino mass, 
since $M$ after perturbation is quite likely to go out of the 3$\sigma$ bound for a 
quasi-degenerate mass spectrum. Therefore, besides optimization of the algorithm 
which includes fast diagonalization of $M$ and 
extracting the neutrino parameters, we use the "generate-and-tune" method: 
we first randomly produce six $\epsilon$ and then choose their phases such that 
the following $\chi^2$-function is minimized: 
\begin{equation}
\chi^2 = \sum_{i} \left(\frac{p_{i}-p_{i}^{0}}{\sigma_{i}}\right)^{2}, 
\end{equation} 
where  $p_i$ are the oscillation parameters which are irrelevant. For example when studying the stability of $\theta_{23}$, we take 
$\delta m^2$, $\Delta m^2$, $\theta_{13}$ and $\theta_{12}$ as irrelevant parameters; 
$p_i^0$ and $\sigma_i$ denote respectively the best-fit values and  $1\sigma$ 
errors of the corresponding parameters from Ref.\ \cite{fit}.
We have checked that the events generated in this way have almost the same distribution
as those generated by the "generate-and-reject" method, but the procedure is 
more efficient and faster. 

One comment should be added here: 
often the mass matrices that are resulting from flavor symmetry models have a 
feature called  
"form-invariance", i.e.\ the eigenvalues do not depend on the mixing angles (infamous tri-bimaximal mixing is one particular example) and our analysis might be irrelevant in this case. However, if breaking terms are added to the mass matrices the form-invariance is lost.

\subsection{Results}

We first look at the correlation of the mixing angles by simply plotting 
$\theta_{23}, \theta_{13}, \theta_{12}$ and $\delta$ after perturbation 
against the original mixing angles $\theta_{23}^{0}, \theta_{13}^{0}, \theta_{12}^{0}$ 
and $\delta^0$ before perturbation. This should give us as feeling on how much the perturbation 
changes the mixing angles. One expects that $\theta_{12}$, being related to the smaller of the two mass-squared differences, will be most unstable. One also suspects $\delta$, being related to phases of various mass matrix elements, to be quite unstable.   
Furthermore, the larger the smallest neutrino mass, the larger the average perturbation. 

The result is shown in Figs.\ \ref{fig:3plots}, \ref{fig:3plots-1}, 
\ref{fig:3plots-2} and \ref{fig:3plots-3} (to illustrate the outcome in a optimal way, we use an 
upper bound $ \sum| \epsilon|<0.04$ instead of $0.2$). 
The plots confirm the expectation.  
When the smallest mass is 0.001 eV, $\theta_{23}$ is
stable, typically deviating from its original value by $\sim1^{\circ}$ (depending  
on the upper bound of $\sum |\epsilon|$). When the smallest mass increases, the
points spread and for 0.1 eV there is hardly any correlation left. 
This conclusion equally applies for $\theta_{13}$, as 
shown in Fig.\ \ref{fig:3plots-1}.  Note that it is the most precisely measured angle, 
and the range of the $y$-axis is much narrower than for $\theta_{23}$. 
However, $\theta_{12}$ and $\delta$ are very unstable even for small masses, 
as can be seen in Figs.\ \ref{fig:3plots-2} and \ref{fig:3plots-3} (note the large range of the axes for the plot with $\delta$). This implies that distinguishing models 
based on precision measurements of $\theta_{12}$ and/or $\delta$ is not a particularly 
reliable method unless corrections are carefully included in the predictions of a model.  
Recall that the plots are for the normal mass ordering. For the inverted ordering, 
$\delta$ and $\theta_{12}$ 
will be uncorrelated with $\delta^0$ and $\theta_{12}^0$ even for the smallest value of 0.001 eV, while $\theta_{23, 13}$ are slightly more uncorrelated (see below).

\newcommand{\widset}{6cm}
\begin{figure}
\centering

\includegraphics[width=\widset{}]{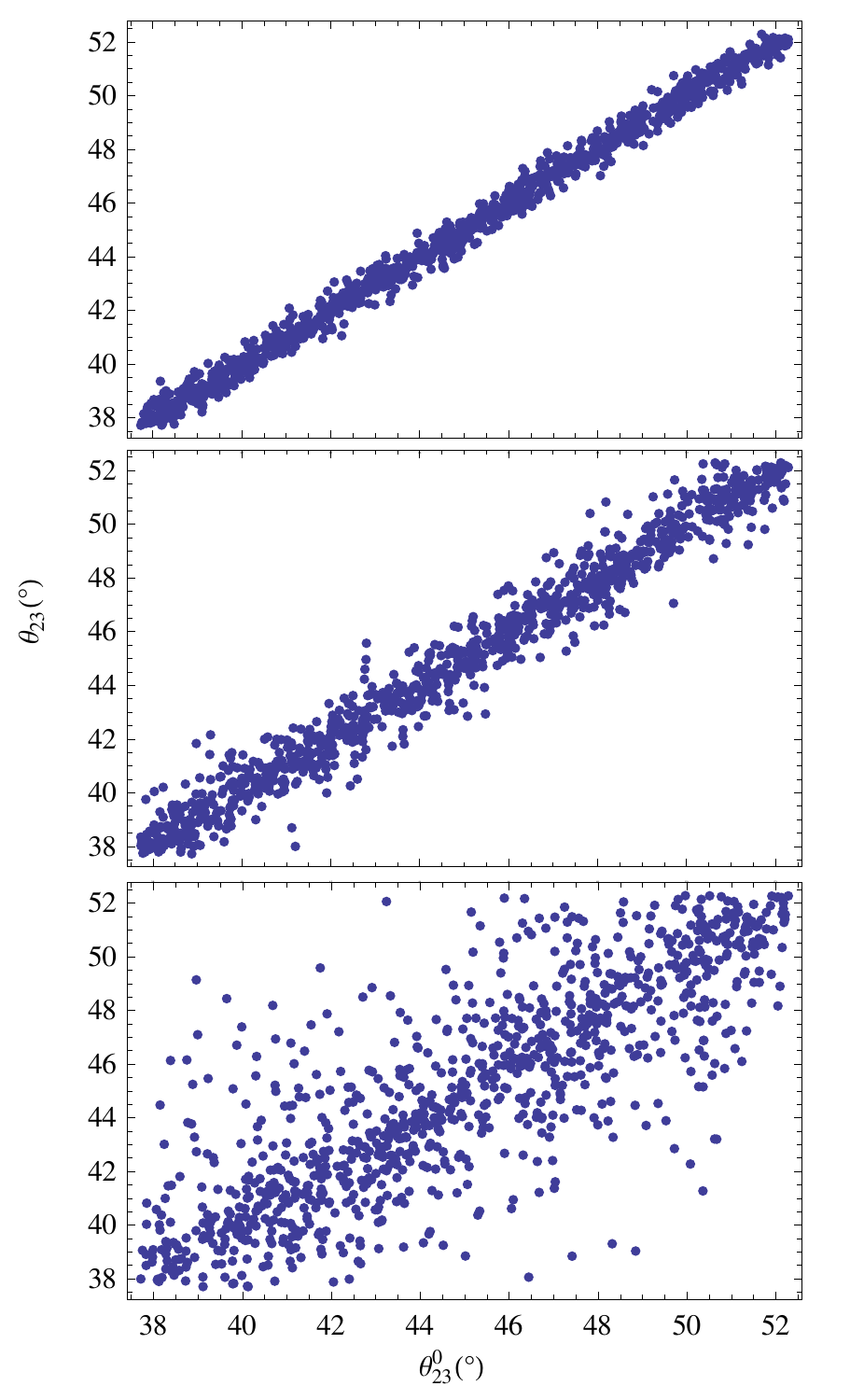}

\protect\caption{\label{fig:3plots}Correlation of $\theta_{23}$ (after perturbation)
with $\theta_{23}^{0}$ (before perturbation). The lightest neutrino
mass is (top to bottom) 0.001, 0.04 and 0.1 eV. }
\end{figure}

\begin{figure}
\centering

\includegraphics[width=\widset{}]{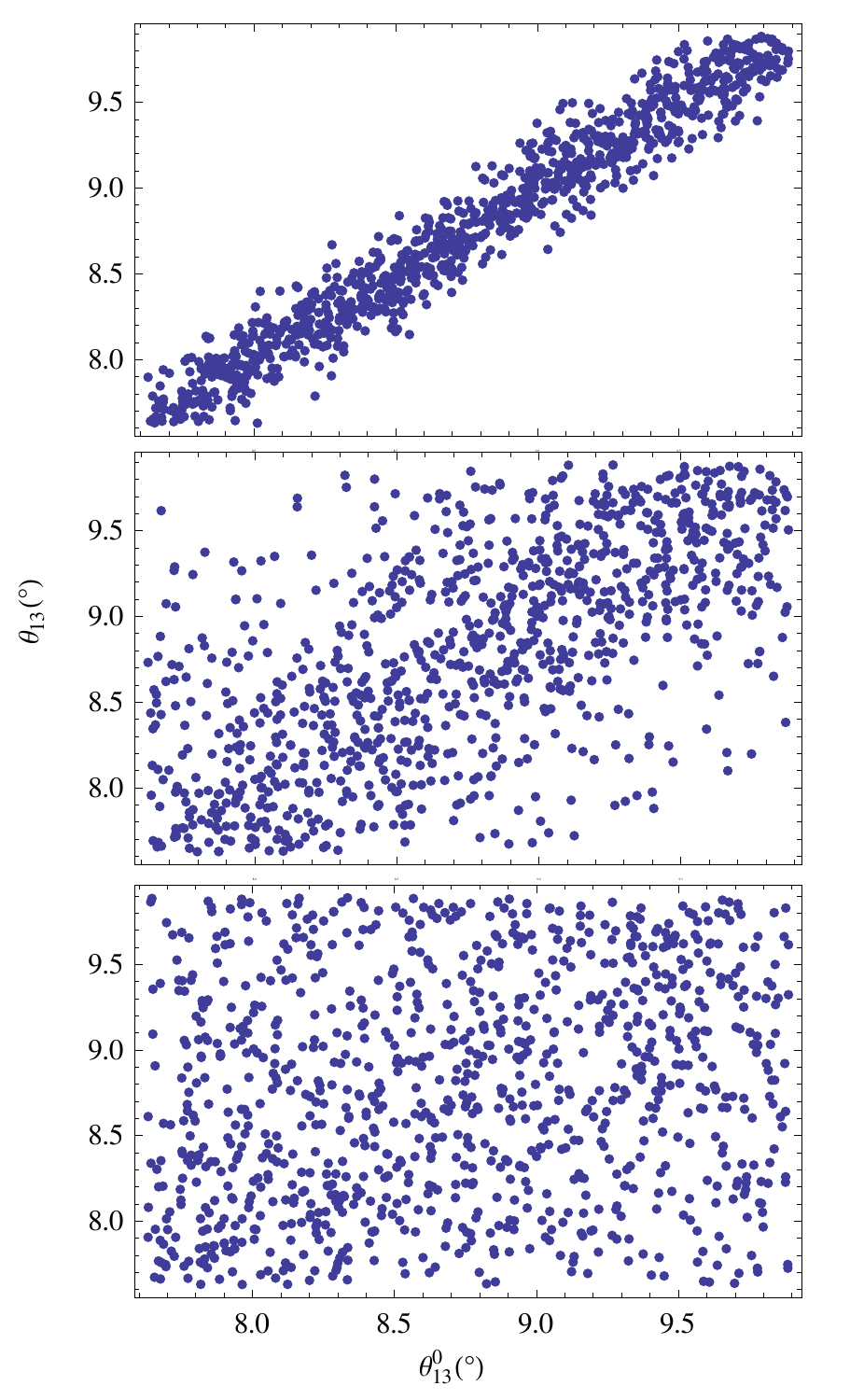}

\protect\caption{\label{fig:3plots-1}
Correlation of $\theta_{13}$ (after perturbation)
with $\theta_{13}^{0}$ (before perturbation). The lightest neutrino
mass is (top to bottom) 0.001, 0.04 and 0.1 eV. 
}
\end{figure}

\begin{figure}
\centering

\includegraphics[width=\widset{}]{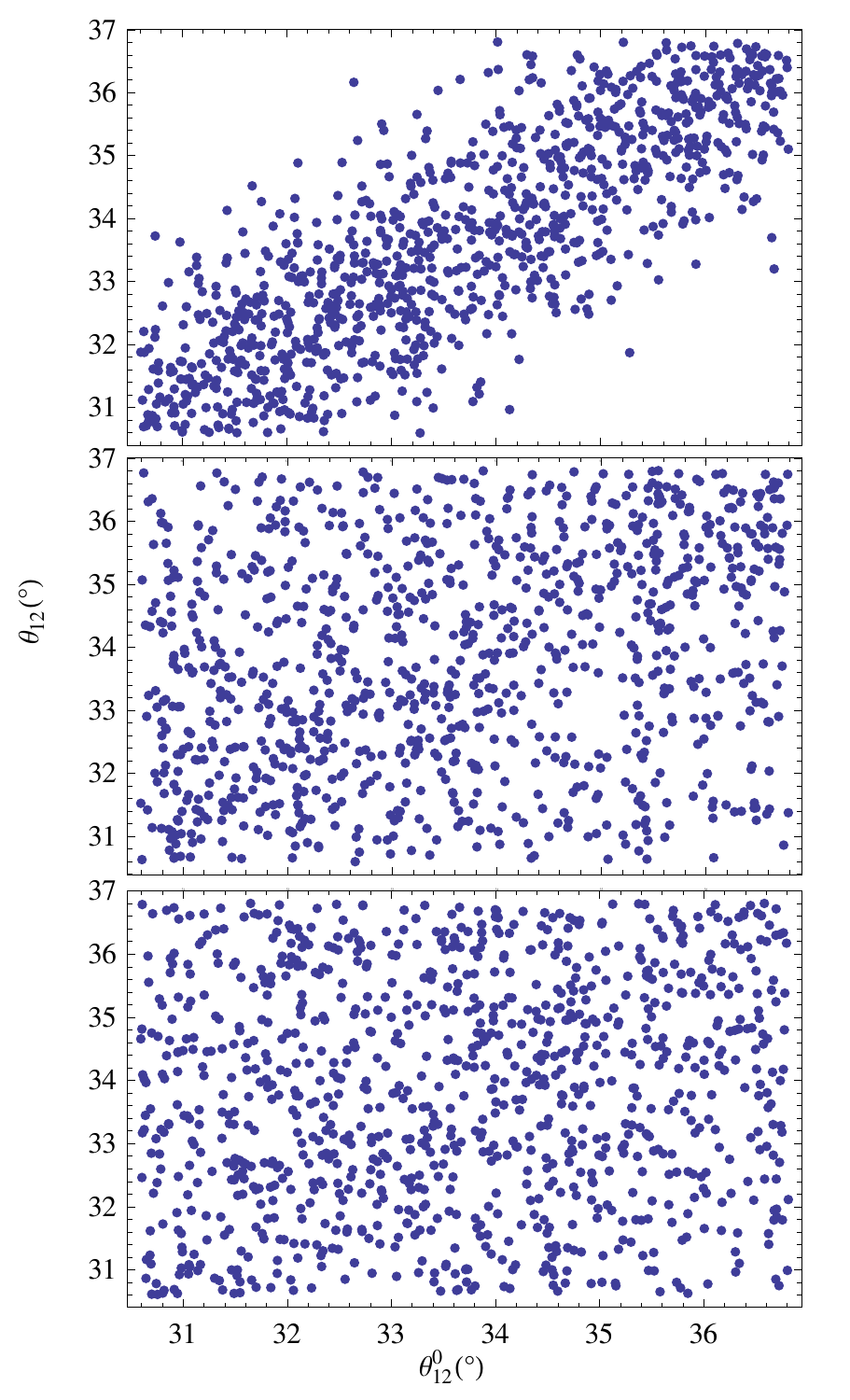} 

\protect\caption{\label{fig:3plots-2}
Correlation of $\theta_{12}$ (after perturbation)
with $\theta_{12}^{0}$ (before perturbation). The lightest neutrino
mass is (top to bottom) 0.001, 0.04 and 0.1 eV.
}
\end{figure}

\begin{figure}
\centering

\includegraphics[width=\widset{}]{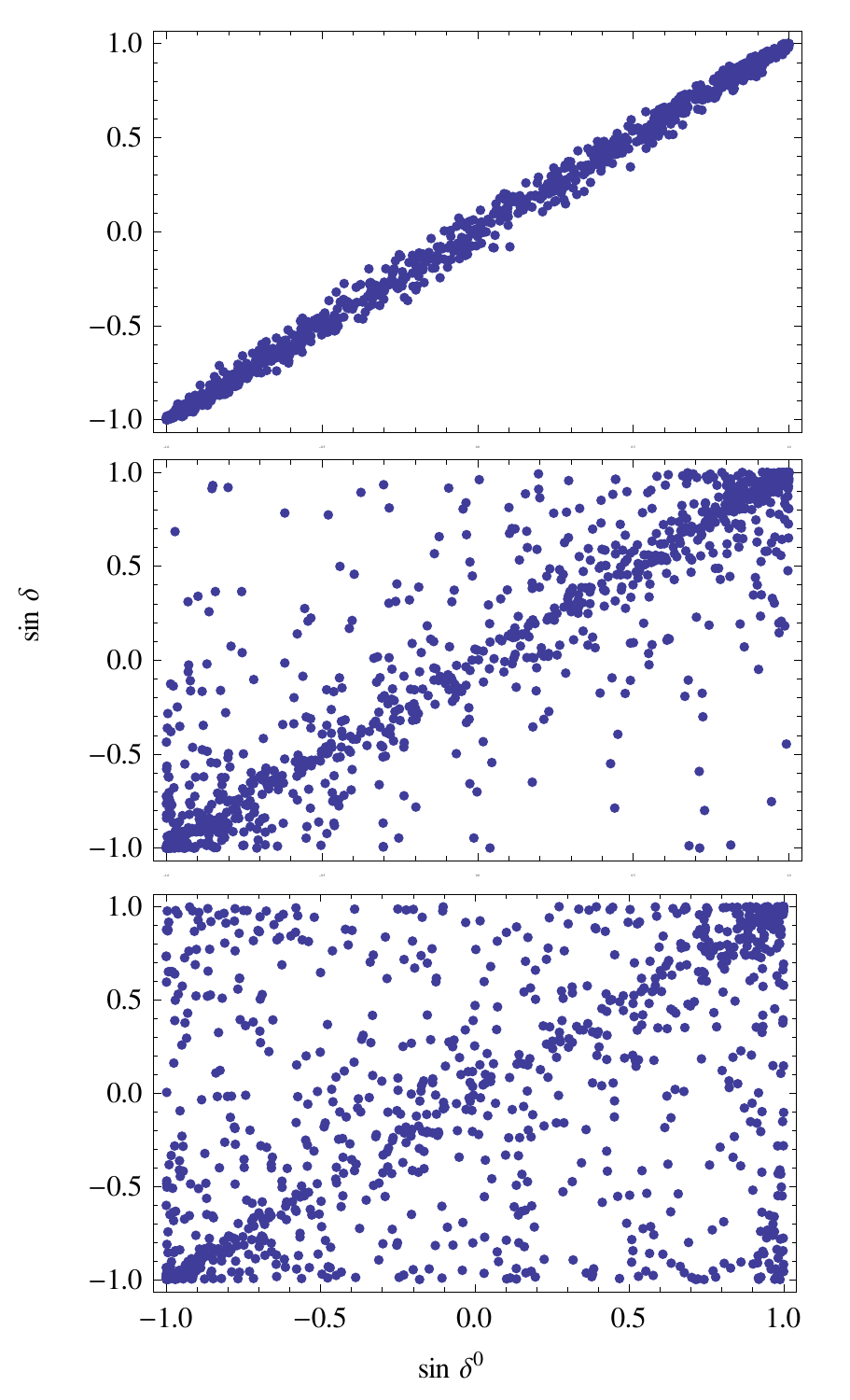} 

\protect\caption{\label{fig:3plots-3}
Correlation of $\sin \delta$ (after perturbation)
with $\sin \delta^{0}$ (before perturbation). The lightest neutrino
mass is (top to bottom) 0.001, 0.04 and 0.1 eV.
}
\end{figure}

\begin{figure}
\centering

\includegraphics[width=7cm,height=5cm]{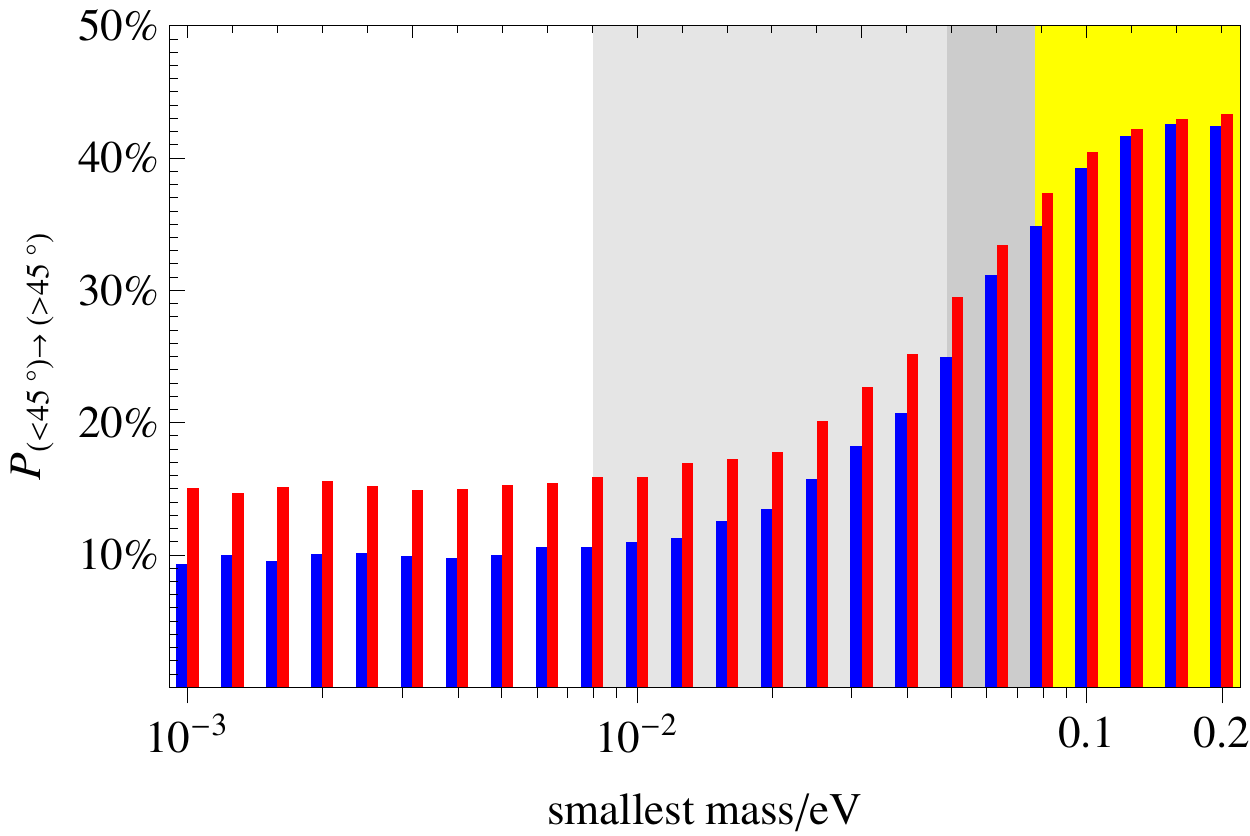}\protect\caption{\label{fig:wrong-octantoct-gen}Percentage of perturbed mass matrices
that give $\theta_{23}$ in the second octant when the unperturbed
mass matrices give $\theta_{23}$ in the first octant (blue for normal mass ordering, red for inverted).  
The light-gray and gray regions start at $\sqrt{\delta m^{2}}\simeq0.008\textrm{ eV}$
and $\sqrt{\Delta m^{2}}\simeq0.05\textrm{ eV}$, respectively, while
the yellow region represents the strongest cosmology constraint on neutrino
masses \cite{planck}. 
}
\end{figure}

\begin{figure}
\centering

\includegraphics[width=7cm,height=5cm]{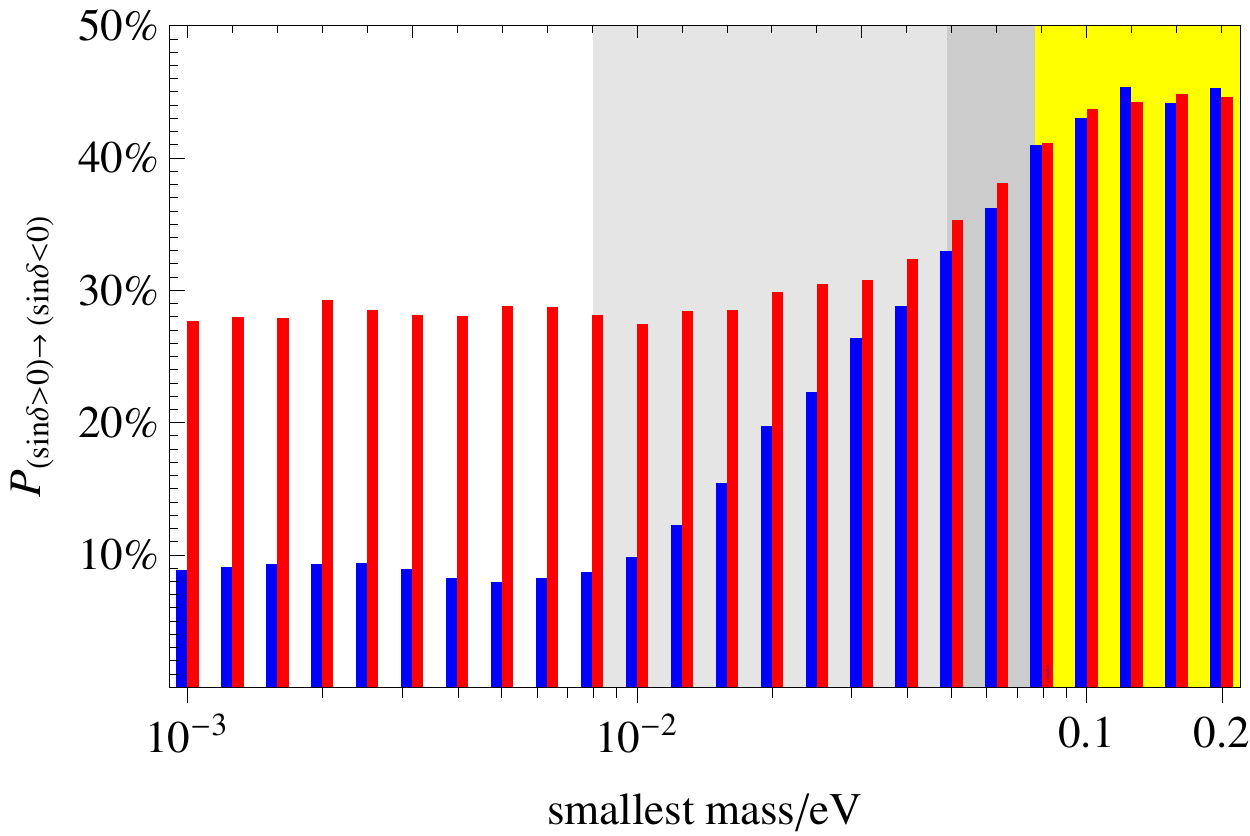}\protect\caption{\label{fig:wrong-s-gen}Percentage of perturbed mass matrices that
give negative $\sin\delta$ when the unperturbed mass matrices give
positive $\sin\delta$ (blue for normal mass ordering and red for inverted). For other details,
see Fig.\ \ref{fig:wrong-octantoct-gen}. }
\end{figure}

\begin{figure}
\centering

\includegraphics[width=7cm,height=5cm]{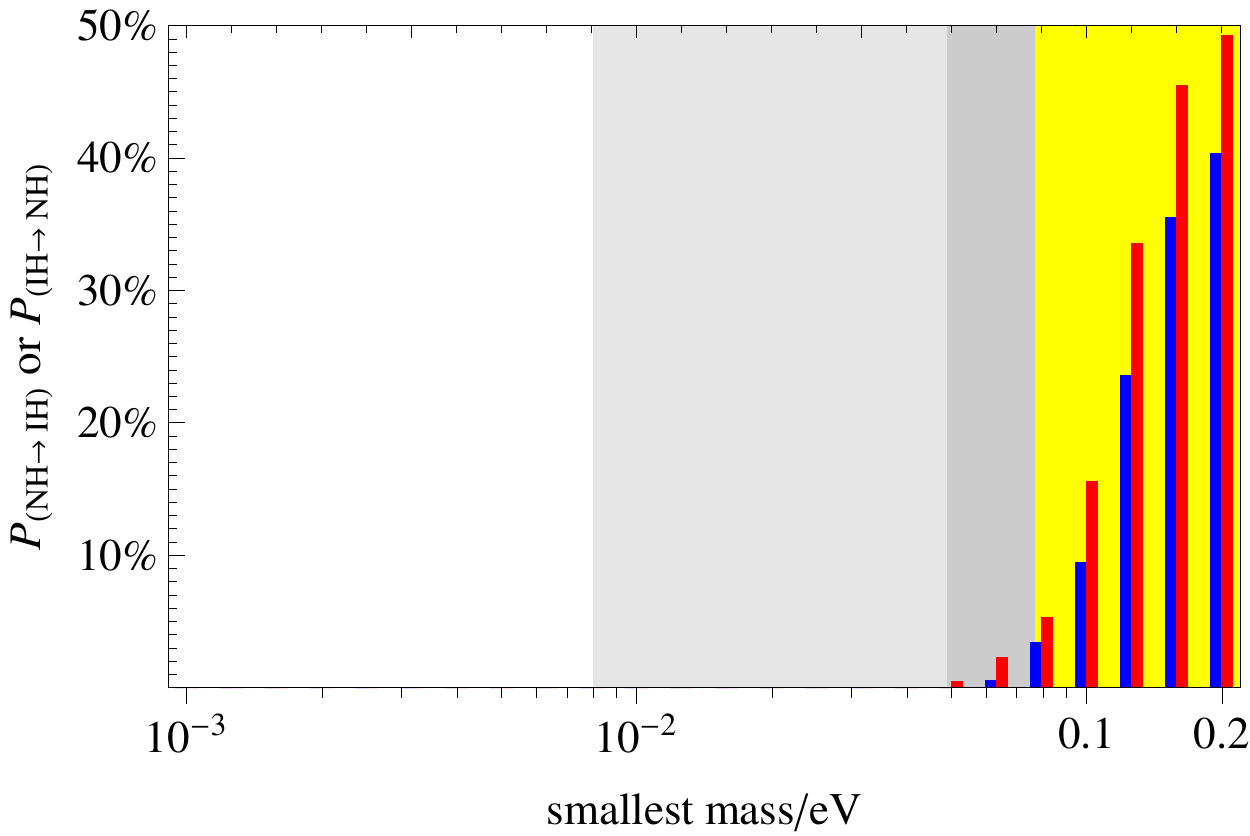}\protect\caption{\label{fig:wrong-h-gen}Percentage of perturbed mass matrices that
give the inverted or normal mass ordering when the unperturbed mass
matrices correspond to the normal (blue) or inverted (red) mass ordering,
respectively. For other details, see Fig.\ \ref{fig:wrong-octantoct-gen}. }
\end{figure}

After these preliminaries, we evaluate now the percentages of the perturbed mass matrices 
that change the octant of $\theta_{23}$, 
the sign of $\sin \delta$ or the mass ordering (we do not consider Majorana phases, as their experimental determination is questionable). The results are shown in 
Figs.\ \ref{fig:wrong-octantoct-gen}, \ref{fig:wrong-s-gen} and 
\ref{fig:wrong-h-gen}. In the plots we indicate two interesting 
mass scales $\sqrt{\delta m^{2}}\simeq0.008\textrm{ eV}$ and $\sqrt{\Delta m^{2}}\simeq0.05\textrm{ eV}$, which will be discussed
in detail later. We also plot the rather strong 95\,\% C.L.\ limit 
on neutrino masses (combining various cosmological data sets) as given by 
Planck \cite{planck}, $\sum m_{i}<0.23\textrm{ eV}$. 

\subsection{Discussion}
Very simple arguments are enough to understand the features of the results. 


In Figs.\ \ref{fig:wrong-octantoct-gen}, \ref{fig:wrong-s-gen} 
\ref{fig:wrong-h-gen} we have indicated two relevant neutrino mass scales, 
$\sqrt{\delta m^{2}}\simeq0.008\textrm{ eV}$ and 
$\sqrt{\Delta m^{2}}\simeq0.05\textrm{ eV}$. 
Let us consider the normal mass ordering. 
Below the mass scale $\sqrt{\delta m^{2}}\simeq0.008\textrm{ eV}$,   
all three mixing angles and the CP phase should be rather stable: 
neither $\delta m^{2}/(m_{1}^{2}+m_{2}^{2})$, associated to 12-mixing, 
nor $\Delta m^{2}/(m_{2}^{2}+m_{3}^{2})$, associated to 13- and 23-mixing,  
are small. 
However, when the smallest mass increases, first  
$\delta m^{2}/(m_{1}^{2}+m_{2}^{2})$ and then $\Delta m^{2}/(m_{2}^{2}+m_{3}^{2})$ 
decrease and become small. Consequently first $\theta_{12}$ and then $\theta_{13,23}$ will become unstable. Increasing the smallest mass further, 
corresponding to more and more quasi-degenerate masses, leads to all 
mixing angles becoming very unstable under perturbations. 

From Fig.\ \ref{fig:3plots-2} we can see that for a smallest mass of $0.04\textrm{ eV}$,
$\theta_{12}$ is unstable (since $\delta m^{2}/(m_{1}^{2}+m_{2}^{2})$ is small) 
as there is no significant correlation 
between $\theta_{12}$ and $\theta_{12}^{0}$. 
In contrast to $\theta_{12}$, $\theta_{13}$ and $\theta_{23}$ are
more stable, as shown in Figs.\ \ref{fig:3plots} and \ref{fig:3plots-1}. 
Increasing the smallest mass, $\theta_{23}$ and $\theta_{13}$ become 
unstable when $\Delta m^{2}/(m_{1}^{2}+m_{3}^{2})$ becomes small, 
which happens when the smallest mass goes beyond 
$\sqrt{\Delta m^{2}}\simeq0.05\textrm{ eV}$. 


More quantitative is Fig.\ \ref{fig:wrong-octantoct-gen}. 
The probability of changing the octant is about 10\,\% for small masses and remains constant until the mass scale $\sqrt{\Delta m^{2}}\simeq0.05\textrm{ eV}$ is reached. Approaching quasi-degenerate masses gives a probability of almost 50\,\% to change the octant, i.e.\ the octant is random and maximally unstable. \\

The predictions are more stable for the normal mass ordering than for the inverted one. 
This conclusion comes from the comparison of the relevant percentages 
in Figs.\ \ref{fig:wrong-octantoct-gen} and \ref{fig:wrong-s-gen}. 
Note that for a vanishing smallest mass we have 
$(m_{1}^{2},m_{2}^{2},m_{3}^{2})\approx(0,\delta m^{2},\Delta m^{2})$ for a normal ordering whereas for the inverted case we have 
$(m_{1}^{2},m_{2}^{2},m_{3}^{2})\approx(\Delta m^{2},\Delta m^{2}+\delta m^{2},0)$. 
Therefore, $\delta m^{2}/(m_{1}^{2}+m_{2}^{2})$ is small from the beginning and of order 
$\delta m^{2}/\Delta m^{2}\approx0.03$. Indeed the probability to change the octant 
starts with about 15\,\% and increases when $\Delta m^{2}/(m_{1}^{2}+m_{3}^{2})$ becomes 
small for smallest masses of 0.05 eV and larger. 
Obviously for quasi-degenerate neutrino masses there will be no difference between the mass orderings. \\


Comparing Fig.\ \ref{fig:wrong-octantoct-gen} and \ref{fig:wrong-s-gen}, 
we can see that between $0.01\textrm{ eV}$ to $0.1\textrm{ eV}$, the probability 
of $\sin\delta$ changing its sign is larger than the probability of 
$\theta_{23}$ changing its octant. This is caused by the fact that phases of eigenvectors 
in a diagonalization procedure are always more sensitive to perturbation than their 
absolute values. 
One might also argue that $\delta$ is related to the Jarlskog invariant $J=$ 
Im$\left( U_{e1} \, U_{\mu 2} \, U_{e2}^\ast \, U_{\mu 1}^\ast \right)$ which
is proportional to $\sin \delta \, \sin \theta_{12} \, \sin \theta_{13}\, \sin \theta_{23} $, 
which means that (for normal ordering) the probability of $\sin \delta$ changing its 
sign should be similar to the probability of $ \theta_{23} $ changing the octant. That is indeed 
what Figs.\ \ref{fig:wrong-octantoct-gen} and \ref{fig:wrong-s-gen} show. 
Note also that $J$ is proportional to 
the imaginary part of $h_{12}\,h_{23}\,h_{13}$, where $h = MM^\dagger$ 
\cite{Branco:2002xf}. For a negligible smallest mass $h$ has a dominating $23$-block in the normal mass ordering, whereas for the inverted ordering it has a democratic structure. 
Taking into account that predictions in 
the inverted ordering are in general less robust motivates to assume that the probability of $\sin\delta\rightarrow-\sin\delta$ is initially much larger 
than for the case of normal ordering. 
Indeed, see Fig.\ \ref{fig:wrong-s-gen}, one starts with almost 30\,\% for small masses. 
Again, for quasi-degenerate masses the sign is essentially random. \\

Interestingly, it is possible to change the mass ordering when perturbations are applied. This requires obviously quasi-degenerate neutrino masses, and 
Fig.\ \ref{fig:wrong-h-gen} shows that for values around 0.1 eV the ordering can change, 
quickly reaching a probability of almost 50\,\%. 
For an inverted ordering the probability is larger and starts for smaller neutrino masses. 
This can be traced to the larger fine-tuning of neutrino masses in the inverted ordering: 
for a smallest neutrino mass of 0.2 eV, we have 
$(m_1, m_2, m_3) = (0.2, 0.200187, 0.206004)$ eV in the normal ordering, but 
$(m_3, m_2, m_1) = (0.2, 0.205822, 0.206004)$ eV in the inverted one (choosing the 
best-values of the mass-squared differences). Therefore, the heaviest and 
next-to-heaviest masses are much closer together in the inverted ordering. After adding 
a perturbation, switching from inverted to normal is thus more likely than the 
other way around.

\fi

\headif

\section{\label{sec:a4}Perturbations on an $A_{4}$ model}

In this section we will see how realistic our findings from the general case treated so far are. 
We apply corrections to a specific flavor symmetry model. 

\subsection{The model}
We consider a model based on the discrete group $A_{4}$, as developed in \cite{Ma:2004zv,Altarelli:2005yp,Barry:2010zk}. 
In the unperturbed limit, the charged leptons are diagonal and the neutrino mass matrix 
is 
\begin{equation}
M_{0}=\left(\begin{array}{ccc}
\frac{2d}{3} & b-\frac{d}{3} & c-\frac{d}{3}\\
b-\frac{d}{3} & c+\frac{2d}{3} & -\frac{d}{3}\\
c-\frac{d}{3} & -\frac{d}{3} & b+\frac{2d}{3}
\end{array}\right).\label{eq:0216}
\end{equation}
The mass-dimension parameters $b$ and $c$ are related to vacuum expectation values (VEVs) of 
$A_4$ singlets $\xi''$ and $\xi'$, respectively. An $A_4$ triplet field $\varphi'$ acquires VEVs in the $(1,1,1)$ direction and governs the parameter $d$: 
\begin{equation}
\langle\xi''\rangle=u_{b}\,,~\thinspace\langle\xi'\rangle=u_{c}\,,~\langle\varphi'\rangle=v'(1,1,1)\,,\label{eq:0322}
\end{equation}
with $b=u_{b}x_{b},c=u_{c}x_{c},d=v'x_{d}$, and $x_{b,c,d}$ are dimensionless parameters. 

Taking $b$, $c$ and $d$ as free parameters, the zeroth order mass
matrix Eq.\ (\ref{eq:0216}) can fit current neutrino data very
well. To obtain the required parameters and to facilitate the perturbation, we minimize the following $\chi^{2}$-function:
\begin{eqnarray}
 &  & \chi^{2}(b,c,d)\equiv\left(\frac{\theta_{23}-\theta_{23}^{0}}{\sigma_{23}}\right)^{2}+\left(\frac{\theta_{12}-\theta_{12}^{0}}{\sigma_{12}}\right)^{2} \nonumber\\
 &  & +\left(\frac{\theta_{13}-\theta_{23}^{0}}{\sigma_{13}}\right)^{2}+\left(\frac{\delta m^{2}-\delta m_{0}^{2}}{\sigma_{\delta m^{2}}}\right)^{2}+ \left(\frac{\Delta m^{2}-\Delta m_{0}^{2}}{\sigma_{\Delta m^{2}}}\right)^{2}\nonumber\\
 &  & +\left(\frac{m-m_{\rm sm}}{\sigma_{m_{\rm sm}}\rightarrow0}\right)^{2},\label{eq:0216-2}
\end{eqnarray}
where the last term is added to fix the smallest mass, $m_{\rm sm}$, which is
implemented by taking $\sigma_{m_{\rm sm }}\rightarrow0$. In practice,
we take $\sigma_{m_{\rm sm}}=m_{\rm sm}/1000$. The parameters
$b$, $c$ and $d$ are, in general, complex numbers. We can remove 
an overall phase so only five degrees of freedom remain. 
The above mass matrix (\ref{eq:0216}) is (partly) form-invariant, the eigenvector to the eigenvalue $b+c$ is always $(1,1,1)^T$, hence $|U_{e2}|^2 = \frac 13$, independent of the 
magnitude of the mass matrix entries. Corrections will destroy this feature. 

As usual in models of this kind, the parameters $b$, $c$ and $d$ have to be somewhat tuned to get the experimental values of $\delta m^{2}$ and $\Delta m^{2}$, which makes it technically difficult to find the minimum of the $\chi^{2}$-function.  
The $\chi^{2}$-fit gives the following conclusions:
\begin{enumerate}
\item The minimal value is non-zero, $\chi_{\rm min}^{2}=3.7$ which implies reasonable 
agreement with current data. The global minima are not unique but discrete, we find that there are four degenerate 
minima with $\chi_{\rm min}^{2}=3.7$;
\item The reason why we cannot have arbitrarily small $\chi^2$ 
is because of $|U_{e2}|^2 = \frac 13$, which forces $\sin^2 \theta_{12}$ to values larger 
than $\frac 13$ ($\sin^2 \theta_{12} = 0.341$, to be precise), while the $1\sigma$-range from global fits is below $\frac 13$. 
The other oscillation parameters can be reproduced to their 
best-fit values at the $\chi^{2}$-minimum, in particular we have \cite{fit} 
$\sin^2 \theta_{23} = 0.437$. Due to the constraint $|U_{\alpha 2}|^2 = \frac 13$, 
one has $\sqrt{2} \, |U_{e3}| \cos \delta \simeq 1/\tan 2 \theta_{23}$, leading to 
$\delta= \pm55.3^{\circ}$; 
\item The degeneracy between the four minima corresponds to  $(\delta,-\alpha_{1,2})\leftrightarrow(\delta,\alpha_{1,2})\leftrightarrow-(\delta,\alpha_{1,2})\leftrightarrow(-\delta,\alpha_{1,2})$. 
Here $\delta$ is the Dirac phase and $\alpha_{1,2}$ are Majorana
phases. For $(\delta,\alpha_{1,2})\leftrightarrow-(\delta,\alpha_{1,2})$ the degeneracy 
is obvious since it means conjugating $M_{0}$. 
Henceforth, we name the four solutions $(+,+)$, $(+,-)$, $(-,+)$ and $(-,-)$ if the signs
of $(\sin\delta,\sin\alpha_{1})$ are $(+,+)$, $(+,-)$, $(-,+)$
and $(-,-)$, respectively. The four discrete minima imply that for a fixed smallest mass,
the model predicts definite CP phases, both Dirac and Majorana. Two 
of the four solutions have positive $\sin\delta$ and two negative; 
\item We do not need to perturb these four zeroth order mass matrices separately, as the $(+,+)$ case is identical to the $(-,-)$ case, and the $(+,-)$ case identical to $(-,+)$; 
\item For $m_{\rm sm}\ls 0.03$ eV (normal ordering) or $0.02$ eV (inverted ordering), 
there is no solution as $\chi_{\rm min}^{2}$
increases rapidly and soon gets out of the $5\sigma$ range. 
For values below, $\chi_{\rm min}^{2}$ is always $3.7$. 
The reason for this is the neutrino mass sum-rule 
$\tilde m_3 - \tilde m_1 = \tilde m_2$ (here $\tilde m_i$ are complex masses, thus including the Majorana phases), which implies \cite{Barry:2010yk}  the relations 
$m_1 \gs \sqrt{\Delta m^2}/\sqrt{3}$ and $m_3 \gs \sqrt{\Delta m^2}/2$, respectively.   
\end{enumerate}

\subsection{Perturbation}

\begin{figure}[t]
\centering

\includegraphics[width=6.5cm]{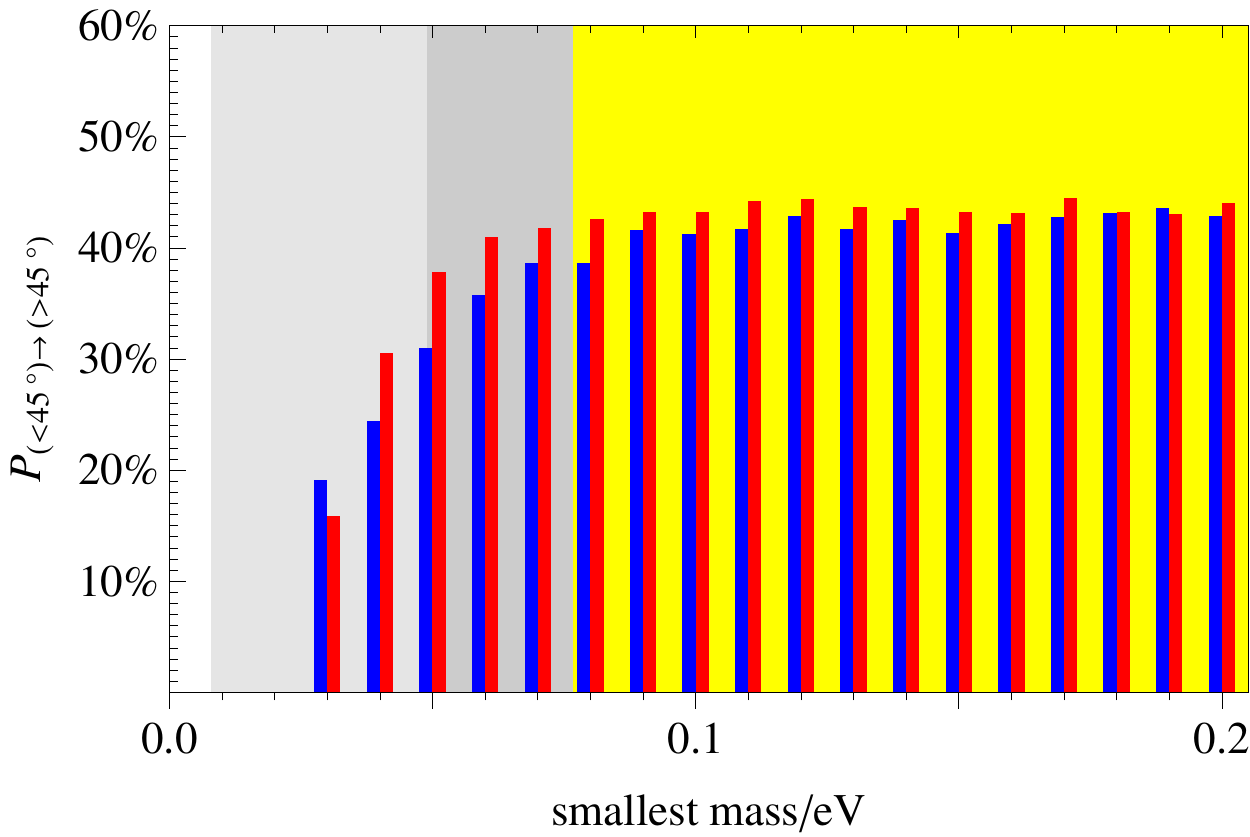}

\protect\caption{\label{fig:wrong-octantoct}Change of octant: 
same as Fig.\ \ref{fig:wrong-octantoct-gen} for the $A_{4}$ model with VEV misalignment. Blue is the normal mass ordering, red  inverted. 
The perturbation is made on the discrete 
solutions $(+,+)$ and $(-,-)$. }
\end{figure}

\begin{figure}
\centering

\includegraphics[width=6.5cm]{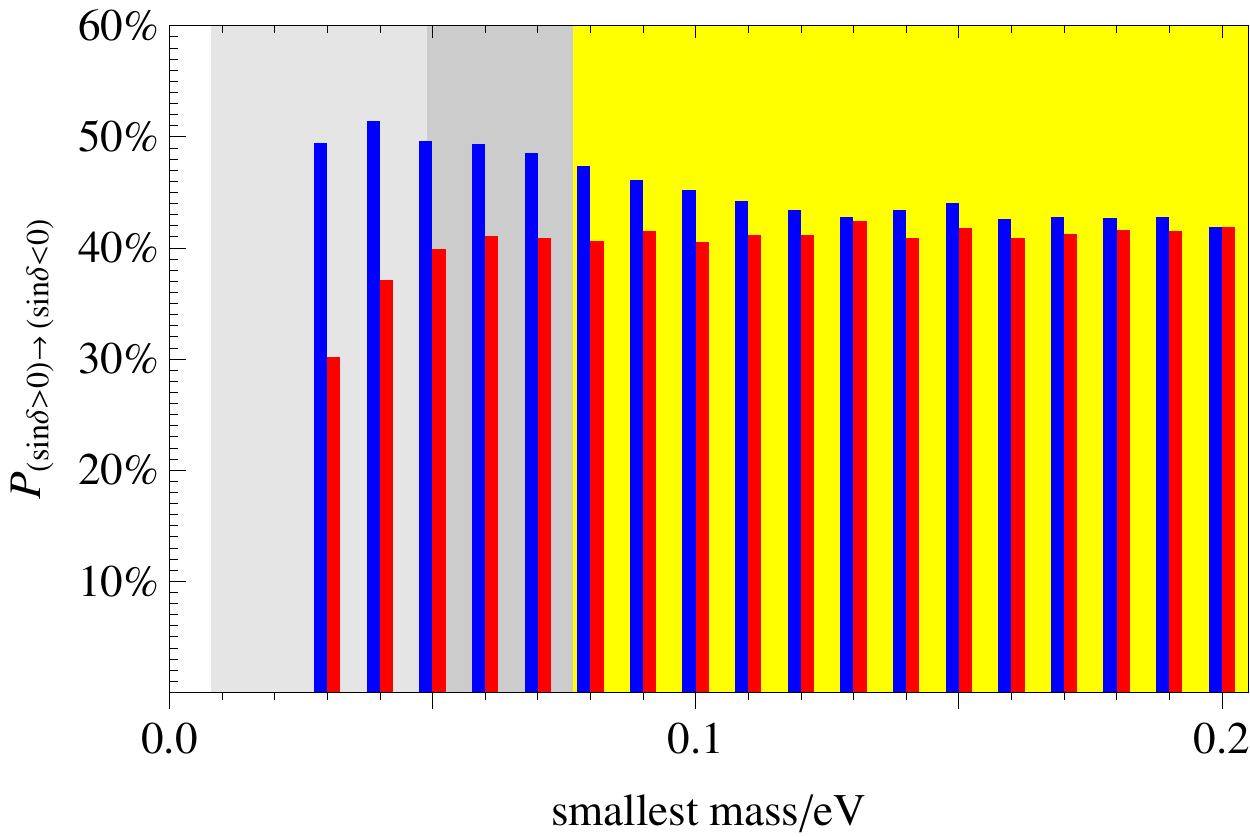}

\protect\caption{\label{fig:flip-the-sign}Change of the sign of $\sin \delta$: same as  Fig.\ 
\ref{fig:wrong-s-gen} for the $A_{4}$ model with VEV misalignment. Blue is the normal mass ordering, red  inverted. 
The perturbation is made on the discrete 
solutions $(+,-)$ and $(-,+)$. 
}
\end{figure}

\begin{figure}
\centering

\includegraphics[width=6.5cm]{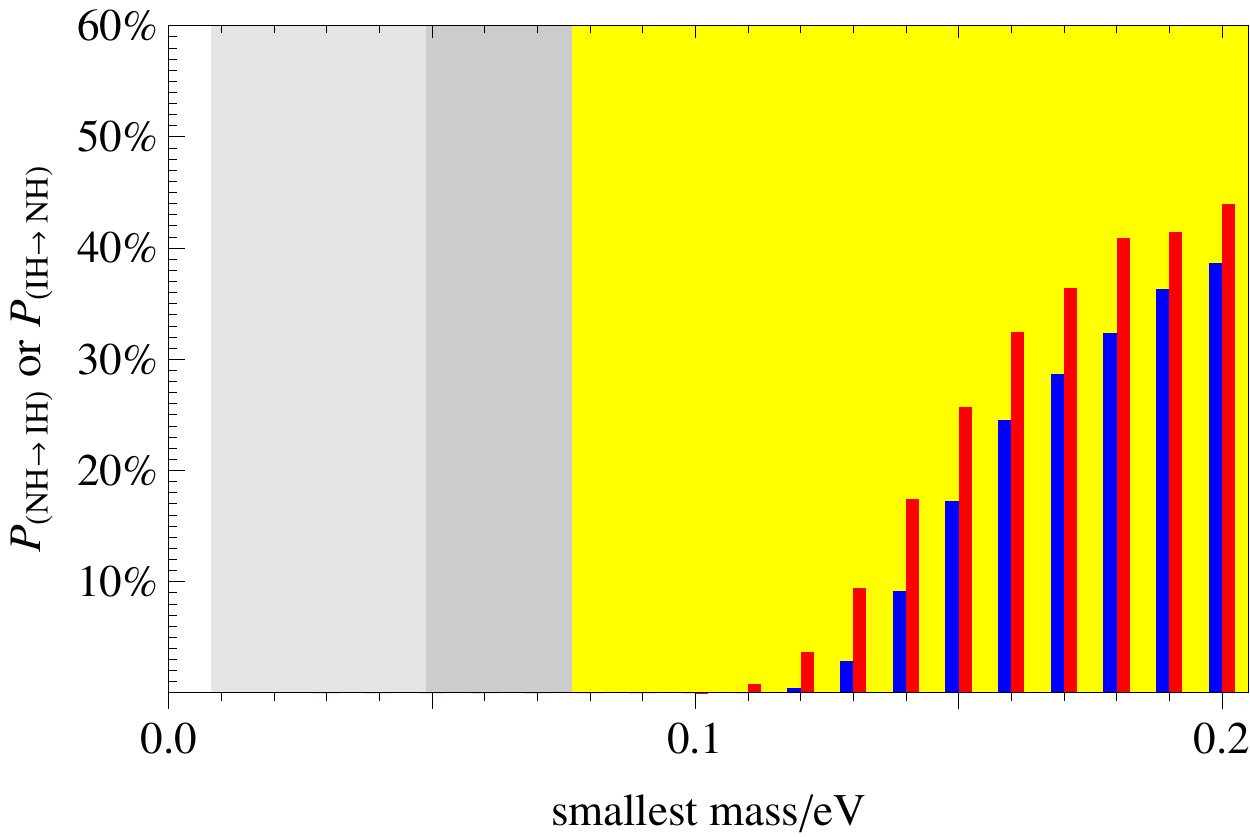}

\protect\caption{\label{fig:hierachy}Change of mass ordering: same as 
Fig.\ \ref{fig:wrong-h-gen} for the $A_{4}$ model with VEV misalignment. Blue is the normal mass ordering, red the inverted. 
The perturbation is made on the discrete 
solutions $(+,-)$ and $(-,+)$.
}
\end{figure}

For simplicity we study the following simple VEV misalignment: 
\begin{eqnarray}\label{eq:0322-03}
\langle\xi''\rangle=u_{b}(1+\epsilon_{1})\,,\thinspace\langle\xi'\rangle=u_{c}(1+\epsilon_{2})\,,
\\
\langle\varphi'\rangle=v'(1+\epsilon_{3},1+\epsilon_{4},1+\epsilon_{5})\,.\label{eq:0322-3}
\end{eqnarray}
As a result, the mass matrix is 
\begin{widetext}
\begin{equation}
M=\left(\begin{array}{ccc}
\frac{2}{3}d\left(1+\epsilon_{3}\right) & b\left(1+\epsilon_{1}\right)-\frac{1}{3}d\left(1+\epsilon_{5}\right) & c\left(1+\epsilon_{2}\right)-\frac{1}{3}d\left(1+\epsilon_{4}\right)\\
b\left(1+\epsilon_{1}\right)-\frac{1}{3}d\left(1+\epsilon_{5}\right) & c\left(1+\epsilon_{2}\right)+\frac{2}{3}d\left(1+\epsilon_{4}\right) & -\frac{1}{3}d\left(1+\epsilon_{3}\right)\\
c\left(1+\epsilon_{2}\right)-\frac{1}{3}d\left(1+\epsilon_{4}\right) & -\frac{1}{3}d\left(1+\epsilon_{3}\right) & b\left(1+\epsilon_{1}\right)+\frac{2}{3}d\left(1+\epsilon_{5}\right)
\end{array}\right).\label{eq:0307-1}
\end{equation}
\end{widetext}
Recall that the original $b,c,d$ are fixed by our initial $\chi^2$ minimization, so 
the various $\epsilon_i$ parameters are indeed required. 
Similar to our study of general perturbations, we randomly generate the $\epsilon$ and study the robustness of the mass matrix, i.e.\ the stability of the octant of $\theta_{23}$, the sign of
$\sin\delta$ and the mass ordering.  
Figs.\ \ref{fig:wrong-octantoct}, \ref{fig:flip-the-sign} and \ref{fig:hierachy} show the result. 
To illustrate our findings, we use $ \sum|\epsilon_i|<0.04$ for Fig.\ \ref{fig:wrong-octantoct} and Fig.\ \ref{fig:flip-the-sign}, but $\sum |\epsilon_i|<0.2$ for Fig.\ \ref{fig:hierachy}; in addition we do not give results for all zeroth order solutions corresponding to the signs of the phases.

As one would expect from the general analysis in Section \ref{sec:gen}, all percentages in the figures increase for increasing smallest mass. The normal mass ordering is somewhat more tuned than the inverted one, as the neutrino mass sum-rule $\tilde m_3 - \tilde m_1 = \tilde m_2$, which holds also after perturbation to good precision, requires more tuned Majorana phases in the normal ordering \cite{Barry:2010yk}. Therefore, the difference between normal and inverted ordering is not as large as in the general case, but the overall structure of the plots in 
Figs.\ \ref{fig:wrong-octantoct}, \ref{fig:flip-the-sign} and \ref{fig:hierachy} is the same as in the general case.   


\fi 

\iftrue

\section{\label{sec:concl}Conclusion}

We have studied the robustness of neutrino mass matrix predictions in the general case and 
within a specific flavor symmetry model. We illustrate the need to include corrections to a mass 
matrix by showing that the octant of $\theta_{23}$, the sign of $\sin \delta$, or even the 
mass ordering can change when perturbations are added. Most of the results are intuitively clear: 
$\theta_{12}$ and $\delta$ are more unstable than $\theta_{13}$ and $\theta_{23}$, thus putting doubt on the 
discriminating power of the solar neutrino mixing angle and the CP phase when corrections are ignored. Predictions from an inverted mass 
ordering are more unstable than the normal one. The larger neutrino 
masses are, the more unstable are the predictions. Going beyond 0.1 eV can even 
change the mass ordering from normal to inverted, quickly reaching a probability of 
50\,\%. 

We have made here conservative assumptions about the perturbation parameters, namely  
multiplicative corrections. Additive corrections are expected to give qualitatively 
similar perturbations, but at least as sizable as the multiplicative 
ones under study here, as they influence small entries 
of the mass matrix more significantly. We have also not considered often and extensively studied 
charged lepton corrections, which are model-dependent and only influence the mixing matrix, 
independent of neutrino mass and mass ordering. At the current stage, we feel that our results 
already illustrate potential issues with the discriminative power of mass matrix predictions when 
perturbations are ignored, but at least illustrate quantitatively 
the potential impact they can have.

\fi 
\begin{acknowledgments}
WR is supported by the Max Planck Society in the project MANITOP, 
XJX by the China Scholarship Council (CSC).
\end{acknowledgments}

\bibliographystyle{apsrev4-1}
\bibliography{ref}

\end{document}